\documentclass[10pt,
               pre,
               showpacs,
               showkeys, 
               twocolumn
              ]{revtex4}
\usepackage{graphicx}      
\usepackage{amsmath,amssymb}

\begin{document}              

\title{Recurrence time analysis, long-term correlations, and extreme events}
\author{Eduardo G. Altmann\footnote{Electronic Address:edugalt@pks.mpg.de}}
\affiliation{Max Planck Institute for the Physics of Complex Systems, N\"othnitzer
  Strasse 38,  01187 Dresden, Germany.} 
\author{Holger Kantz}
\affiliation{Max Planck Institute for the Physics of Complex Systems, N\"othnitzer
  Strasse 38,  01187 Dresden, Germany.} 

\date{\today}

\begin{abstract}

\noindent The
recurrence times between extreme events have been the central point of
statistical analyses in many different areas of science.
Simultaneously, the Poincar\'e recurrence time has been
extensively used to characterize nonlinear dynamical systems. 
We compare the main properties of these statistical methods
pointing out their consequences for the recurrence analysis performed
in time series.  In particular, we analyze the dependence of the mean recurrence time
 and of the recurrence time statistics on the
probability density function, on the interval whereto the recurrences
are observed, and on the temporal correlations of time series. In the
case of long-term correlations, we verify the validity of the
stretched exponential distribution, which is uniquely defined by the
exponent~$\gamma$, at the same time showing that it is restricted to
the class of linear long-term correlated processes. 
Simple transformations are able to modify the
correlations of time series leading to stretched exponentials recurrence time
statistics with different $\gamma$, 
which shows a lack of invariance under the change of observables.
\end{abstract}


\maketitle


\section{\label{sec.I} Introduction}

Recurrence time analysis is a powerful tool to characterize temporal
properties of well defined events~\cite{bunde,bunde.prl}. It has been recently
extensively performed in a rich variety of experimental time series:
records of the climate~\cite{bunde2,santhanam,alley}, seismic
activities~\cite{earthquakes}, solar flares~\cite{boffetta.prl},
spikes in neurons~\cite{joern}, turbulence in magnetic confined
plasma~\cite{murilo.plasma} and stock market indices~\cite{murilo.bolsa}.
Calculated essentially in the same way, these analyses receive
different names: waiting time distribution, interocurrence time
statistics, distribution of interspike intervals, distribution of
laminar phases, etc.  
Through a unified perspective, we discuss the
main properties of these statistical methods, which allows us to
reinterpret and specify many previous results. By a careful discussion of the relevance of the probability
density function (PDF) of the time series data we can easily
understand and, in a particular case, reject results on earthquake
statistics (Sec.~\ref{sec.II}). In the case of
long-term correlated linear time series we obtain a closed expression
for the stretched exponential distribution of recurrence times which is valid for
different recurrence intervals. We show also the lack of invariance of
the long-term correlations of the time series under transformations
that simulate the choice of different observables of the system (Sec.~\ref{sec.III}).
Before reporting these results we start with a proper definition of time
series recurrence times, we compare it to the Poincar\'e recurrences and we
discuss briefly some important recent applications of the recurrence time
analysis. 

\subsection{Recurrence times}\label{ssec.define}\label{ssec.poincare}

\noindent{\sl Time series recurrences}\\
Assuming a time series point of view, 
in this paper we study the statistical properties of the {\em
recurrence time}~$T$. Given a time series $\{x_n\}, n=1,\ldots,N$, and 
having defined a recurrence interval $I$ as a
subset of the data range, then the $i$th recurrence time $T_i$
is the time interval $\Delta n$ between the $i$th and the $i+1$st visit of
a time series point in $I$. The recurrence time statistics (RTS) is obtained as
the distribution $P(T)$ of the sequence of recurrence times~$T_i$.

\begin{figure}[!ht]
\centerline{\includegraphics[width=\columnwidth]{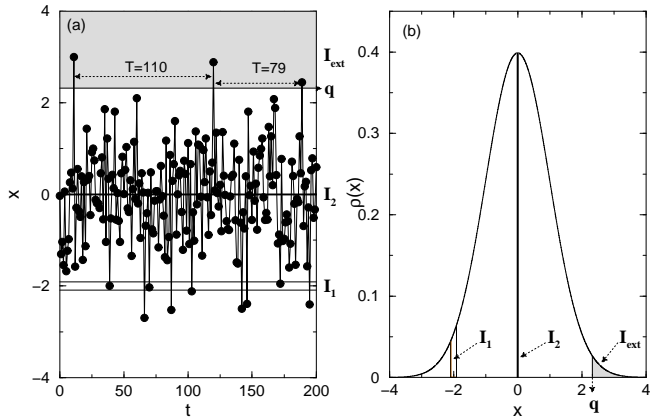}}
\caption{Gaussian distributed time series with recurrence intervals with
  measure~$\mu(I)=0.01$, what implies $\langle T \rangle = 100$. One extreme
  interval~$I_{ext}(q=2.3263)=[2.3263,\infty[$ and two recurrence intervals
  $I_1(-2,0.0922)=[-1.9078,-2.0922]$, $I_2(0,0.0125)=[-0.0125,0.0125]$
  illustrated 
  in the (a) time series and (b) probability density function.}
\label{fig.gaussian}
\end{figure}

Evidently, the sequence of recurrence
times generated this way depends sensitively on the choice of $I$,
which in fact will be one prominent issue of this paper. 
While for the recurrence of extreme events the recurrence
interval is defined by the points above a threshold~\cite{bunde}
\begin{equation}\label{eq.q}
I_{ext}(q)=[q,\infty[\;,
\end{equation}
in a more general way it may be defined around a position~$X_c$ with a
semi-width~$\delta$~\cite{murilo.plasma,altmann} 
\begin{equation}\label{eq.xc}
I(X_c,\delta)=[X_c-\delta,X_c+\delta]\;.
\end{equation}
Both kinds of intervals are illustrated in Fig.~\ref{fig.gaussian}.\\

\noindent{\sl Poincar\'e recurrences}\\
In dynamical systems' theory, another concept of recurrence, known as 
Poincar\'e recurrences, plays a central role. Given a closed
Hamiltonian system with ergodicity on the energy shell, the famous
Poincar\'e recurrence theorem asserts that almost all trajectories (except
for a set of zero measure) started inside some subset ${\cal V}$ of the 
phase space $\Gamma$  will return to it infinitely many times.
In the limit of vanishing volume of this subset, the time between
consecutive recurrences is the Poincar\'e recurrence time.
Despite the well known debates about the foundations of statistical
mechanics (Zermello paradox) these ideas motivated throughout the years also
mathematical studies~\cite{kac} and, more recently, applications of recurrence
analysis to many different dynamical systems (see Ref.~\cite{zas.pr} and
references therein). 

Surprisingly, as far as we know, no
connection between the two recurrence approaches described above 
were made until now. The most evident way to establish this 
relationship is to define an observable $x=x(\vec \gamma)$, when $\vec
\gamma(t)\in\Gamma$ is the trajectory in phase space of the
Hamiltonian system. The recurrence volume ${\cal V}$ is mapped to an
interval $I_{\cal V}$ on the real axis by the observation function
$x(\vec \gamma)$. However, the sequence of recurrence times of the series
$x_n:=x(\vec\gamma(t=n\Delta t))$ with respect to $I_{\cal V}$ is generally
{\sl not} identical to the sequence of Poincar\'e 
recurrences of $\vec\gamma(t)$
with respect of ${\cal V}$, since there is usually a large set
$\bar{\cal V}$ which also maps to $I_{\cal V}$ due to the
non-invertibility of $x(\vec\gamma)$. Moreover,
generally $I_{\cal V}$ will be of the kind of $I(X_c,\delta)$ rather
than $I_{ext}(q)$. 

However, as we will show in this paper, the
analogy with Poincar\'e recurrences motivates issues related to the
recurrence times of extreme events which will reveal fundamental
insight into their properties. Two main results will be the lack of
invariance of the RTS under change of the
observable and the fact that long-term correlations are {\sl not}
fully characterized by the autocorrelation function.

\subsection{Earthquakes and SOC models}\label{ssec.soc}

The recurrence time between extreme events was recently used in the analysis of different
experimental time
series~\cite{bunde2,santhanam,alley,joern,boffetta.prl,earthquakes}. One
of the most important examples of this analysis, which is going to be
discussed later in this paper, is the study of the waiting time
between earthquakes or avalanches in models exhibiting self-organized
criticality (SOC).  The idea of studying recurrences in SOC started
with the first connections between SOC and
earthquakes~\cite{sornette}. More recently, the investigation of
seismic catalogs of different regions of the globe indicate the
existence of a universal distribution of recurrence times between big
earthquakes~\cite{earthquakes}, which may be roughly described as a
power-law distribution
\begin{equation}\label{eq.powerlaw}
P(T) \propto T^{-\alpha},
\end{equation}
followed by a faster decay. Simple SOC models have a 
Poisson (exponential) distribution of recurrence times, what was used as argument against the
use of SOC to model not only earthquakes~\cite{yang.prl} 
but also
(and originally) solar flares~\cite{boffetta.prl}. However, non-Poissonian
distributions are obtained in more sophisticated SOC
models~\cite{sanchez.prl,christiansen}, what keeps open the debate  over the 
use of SOC in these fields, with the RTS as one of its central ingredients.

\subsection{Long-term correlations and recurrence times}\label{ssec.bunde}
If time series data $\{x_n\}$ are  exponentially (short range)
correlated, the RTS is well known to be  Poissonian, i.e., exponential for
all~$T$, independent of the choice of $I$ (in the limit of small interval~$\mu(I) \rightarrow
0$)~\cite{altmann}. The same result applies to Poincar\'e recurrences (including independence
of ${\cal V}$) if the underlying dynamics is hyperbolic, i.e., in well defined mathematical way fully
chaotic~\cite{hirata}. Also in this case, correlations decay fast. Hence, for
systems with an exponential decay of correlations, details of defining
recurrence times and further details of the system are irrelevant; instead
there exists a unique RTS.

Many time series data have been found to be long-term correlated,
i.e., their mean autocorrelation time
diverges~\cite{bunde2,santhanam,kantelhardt}. Typically, this situation is
characterized in the time series $\{x_n\}$
(assuming $\langle x_n \rangle=0$) by the 
exponent~$0< \gamma_c <1$ of the power-law decay of the
autocorrelation function as a function of the time~$s$ 
\begin{equation}\label{eq.correlation}
C_x(s) = \langle x_i x_{i+s} \rangle = \frac{1}{N-s} \sum_{i=1}^{N-s} x_i
x_{i+s} \sim s^{-\gamma_c}\;.
\end{equation}

In a recent paper~\cite{bunde}, Bunde et al.\ analyzed the effect of
long-term correlations on the return periods of extreme events, i.e.,
of recurrence times obtained using recurrence intervals of type (\ref{eq.q}).  The main results of Ref.~\cite{bunde,bunde.prl}
for long-term correlated time series
can be summarized
by the following three points. While the first was obtained
considering statistical arguments, the two others were based on
numerical simulations.\\
{\em(i)} The mean recurrence time is equal to the inverse of
  the fraction of 
  extreme points in the series 
$$\langle T \rangle = \frac{N_{total}}{N_{extreme}}.$$
{\em(ii)} The statistics of~$T$ follows a stretched exponential
\begin{equation}\label{eq.bunde}
\ln P(T) \propto -(T/\langle T\rangle)^\gamma\;,
\end{equation}
where $\gamma=\gamma_c$ is identical to the correlation exponent in Eq.~(\ref{eq.correlation}). \\
{\em(iii)} The series of recurrence times is long-term correlated with an
  exponent~$\gamma_T$ close to~$\gamma_c$. 

Statement {\em(i)} is the time series analogous of Kac's Lemma (Sec.~\ref{sec.II}),
statement {\em(ii)} will be verified carefully (Sec.~\ref{ssec.numerical}) once we have
established the full functional form of the stretched exponential
(\ref{eq.bunde}), and statement {\em(iii)} seems not to be generally valid (Sec.~\ref{ssec.seriesT}). 

Even if one might argue that based on Ref.~\cite{bunde} 
the validity of Eq.~(\ref{eq.bunde}) 
is established only for the class of the model data
chosen there, the reproduction of these findings for empirical
data~\cite{bunde2,santhanam} suggests
some generality of the stretched exponential distribution. Here, the link to
Poincar\'e recurrences shows the opposite: 
Hamiltonian systems with mixed phase
space are long-term correlated and 
show power-law tails in the statistics of Poincar\'e recurrence
times~\cite{zas.pr}. 
In this case, the long-term  
correlations are originated by the stickiness of chaotic trajectories near the
border of integrable islands. They cause a kind of intermittent
dynamics and manifest themselves in complicated higher-order temporal
correlations. 
In fact, the temporal properties of typical data are 
{\em not} fully specified by the autocorrelation function,
Eq.~(\ref{eq.correlation}), what explains why there cannot be a unique
RTS for long-term correlated data. Connections between the
 long-term correlation exponent~$\gamma_c$ and the RTS have to be
established independently in every class of long-term correlated dynamical
systems, as was done for Hamiltonian systems with mixed phase
space~\cite{zas.pr} and 
fractal renewal point processes~\cite{thurner}. We argue in
Sec.~\ref{ssec.numerical} that the results of Ref.~\cite{bunde} described
above are valid for
long-term correlated {\em linear} time series~\cite{bunde.prl}. In this paper
we propose a closed expression for the RTS of time series
of this class, which is valid for recurrence intervals of both types~(\ref{eq.q}) and~(\ref{eq.xc}).   


\section{Mean recurrence time}\label{sec.II}
The mean recurrence time 
$$\langle T \rangle \equiv \lim_{N_e \rightarrow \infty} \frac{1}{N_e} \sum_1^{N_e} T_i,$$ 
is a direct result of the choice of the recurrence interval. In area
preserving dynamical systems Kac's lemma~\cite{kac} states that the
inverse of $\langle T \rangle$ is equal to the ergodic measure of the  
recurrence interval~$\mu(I)$. In the case of stationary time series, as
illustrated in Fig.~\ref{fig.gaussian}, an equivalent result is obtained from
the normalized PDF $\rho(x)$, 
\begin{equation}\label{eq.kac}
\frac{\Delta t}{\langle T \rangle} = \mu(I) \equiv \int_{x\in I} \rho(x)dx,
\end{equation}
where~$\Delta t$ 
is the sampling rate used to record the time series\footnote{When
  there is no such parameter, as in the series of earthquakes, the time
  scale is defined by the total number of events and the total recording
  time.}. 
This is the most important constraint to the statistics of 
recurrence times.  In the time series analysis this measure is estimated as
the fraction of valid events (points inside the  
recurrence interval)~$\mu(I)=N_{events}/N$. Intuitively,
relation~(\ref{eq.kac}) states simply that the 
total observation time~$t$ is given by 
$$t=N \Delta t = N_{events} \langle T  \rangle.$$ 

Besides the RTS $P(T)$ the PDF $\rho(x)$ of the series of points
itself is typically used to characterize the time series.  Contrary to
other time series analyses (as the detrended fluctuation analysis
discussed below), the RTS is independent of the PDF. In particular, it
is irrelevant whether the second moment of the PDF is finite. A time
series with a well behaved (Gaussian) PDF can have either exponential
or power-law RTS~\footnote{Take for instance the analysis made in
Sec.~\ref{ssec.numerical} which will give the desirable RTS if we
choose~$\gamma_c=1$ or~$\gamma_c\rightarrow0$ respectively.}. Reversely, a time
series with fat tails in the PDF can lead to a RTS that might be
Poisson or power-law~\footnote{These are obtained if we apply the
transformation~(\ref{eq.y}) below again to uncorrelated~$(\gamma=1)$
or correlated~$(\gamma=0)$ time series respectively.}. The reason for
this is simple: the RTS depends on the sequence of the time series
points and changes under their temporal rearrangement, which does not
change the PDF of the data. While the RTS
is independent of the PDF of the series, the opposite happens to the
mean recurrence time~$\langle T \rangle$. Once the recurrence interval
is defined, whether by relation~(\ref{eq.q}),~(\ref{eq.xc}) or by any
other possible definition, the PDF~$\rho(x)$ provides~$\langle T
\rangle$ through relation~(\ref{eq.kac}).

These two apparently trivial observations, i.e., independence of the RTS and
dependence of $\langle T \rangle$ on the PDF, shed new light 
on previous results. In what follows, we exemplify these points in the
analysis of recurrence times between
earthquakes, already mentioned in Sec.~\ref{ssec.soc}. Despite (or
because of) the  
complexity  of this field it has an important simplicity: the
Gutenberg-Richter law  
\begin{equation}\label{eq.gr}
\rho(M) \propto e^{-b \;\ln(10) M}\;,
\end{equation}
where $A,b$ are constants and $M$ is the magnitude of the earthquake, which is
proportional to the logarithm of the released energy. The
constant~$b$ is almost the same for different parts of the world
and the empirical law~(\ref{eq.gr}) is valid for~$2 \leq M \lesssim
7.5$. From our perspective this means that the PDF of the time 
series of seismic activity is given\footnote{This assumption is not
  completely precise since in 
  order to use extreme intervals (defined by Eq.~(\ref{eq.q})) it is 
  necessary to know the PDF in the limit~$M\rightarrow\infty$. In  the
  case of earthquakes it is well known, from general energy considerations,
  that a faster decay of the Gutenberg-Richter law is necessary
  asymptotically. In order to obtain a sufficient statistics in the analyses
  of experimental data, the choice of~$q$  in Eq.~(\ref{eq.q}) is usually
  considerably smaller than~$7.5$ and 
  the influence of the unknown asymptotic of the PDF becomes negligible.}.

The mean recurrence time between earthquakes of a given magnitude~$M$ is
obtained inserting the PDF
given by~(\ref{eq.gr}) in relation~(\ref{eq.kac}), and using 
the interval of the type~(\ref{eq.xc}) with~$X_c=M$,
\begin{equation}\label{eq.sornette2}
\langle T \rangle(M) = T_0 e^{b\; ln(10) M} \;,
\end{equation}
where $T_0 \propto \frac{b \;ln(10)}{(1-e^{-b\;ln(10) \delta})}$.
This relation is equivalent 
to the one obtained previously through a ``mean-field
approach''~\cite{sornette}. In Ref.~\cite{sanchez.prl} it is noted the
``remarkable'' scaling of~$\langle T \rangle(M\ge M_c) \propto 10^{b
  M_c}$, which
  is nothing else than a consequence of relation~(\ref{eq.kac}) when intervals of the
  type~(\ref{eq.q}) are used with~$q=M_c$. 

So far, the relation between~$\langle T \rangle$ and the PDF was used
to show that the mean waiting time between earthquakes is directly
related to the Gutenberg-Richter law, but has nothing to do with
temporal correlations between earthquakes. On the other hand, the RTS
obtained from earthquakes records~\cite{earthquakes} is an independent
result that can be used as a test for the dynamical models of
earthquakes. Recently, it was suggested that in SOC models the
sequence of avalanches is uncorrelated~\cite{boffetta.prl,yang.prl}
(see~\cite{joern2} for a counterexample) and should thus be
discarded. The solution of this debate is beyond the scope of this
paper. Nevertheless, we note that, as a consequence of the
unrelatedness of~$\rho(x)$ and $P(T)$, shuffling data of whatever
distribution randomly (as was done for the time series of seismic activity in
Ref.~\cite{yang.prl}) trivially implies~$P(T)$ of being exponential,
also for finite recurrence intervals~\cite{altmann}.


\section{\label{sec.III} Statistics of recurrence times}

\subsection{Closed expression of the stretched exponential distribution}\label{ssec.closed}

We generalize the distribution proposed in Ref.~\cite{bunde} for the RTS of
long-term correlated time series. Motivated by result {\em (ii)}, mentioned in
Sec.~\ref{ssec.bunde},  suppose that the stretched exponential
distribution  
\begin{equation}\label{eq.stretched1}
P_\gamma(T)=a e^{-(bT)^\gamma}
\end{equation}
is valid for all recurrence times~$T \in~]0,\infty[$. This is
    actually a stronger assumption than Eq.~(\ref{eq.bunde}). As
    any RTS, Eq.~(\ref{eq.stretched1})
must satisfy the following two conditions: normalization 
%
$$\int_0^\infty P(T) dT=1,$$
%
and the analogous of Kac's lemma~(\ref{eq.kac})
\begin{equation*}\label{eq.cond2}
\langle T \rangle \equiv \int_0^\infty TP(T) dT=\frac{1}{\mu(I)}.
\end{equation*}
Imposing these two conditions to the distribution~(\ref{eq.stretched1}), it is
possible 
to express the constants $a$ and $b$ as functions of $\gamma$
and $\mu(I)$. Further simplification is obtained performing the 
following transformation of variables $\tau=\frac{T}{<T>}=\mu(I)T$, i.e.,
counting the time in units of the mean recurrence time. The
complete stretched exponential distribution for recurrence times is
then written as 

\begin{equation}\label{eq.stretched}
p_{\gamma}(\tau) = a_\gamma e^{-(b_\gamma \;\tau)^\gamma},
\mbox{ with} \left\{ \begin{array}{ll} 
a_\gamma=&b_\gamma \; \frac{\gamma}{\Gamma(1/\gamma)},\\ \\
  b_\gamma=&\frac{(2^{1/\gamma})^2 \Gamma (\frac{2+\gamma}{2\gamma})}{2
    \sqrt{\pi}},
\end{array}
\right.
\end{equation}
and depends exclusively on the exponent~$\gamma$.
\begin{figure}[!ht]
\centerline{\includegraphics[width=\columnwidth]{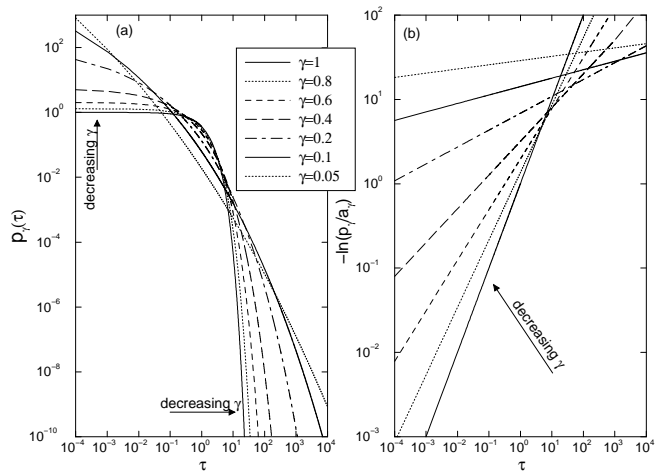}}
\caption{The stretched exponential
  distribution~(\ref{eq.stretched}) for different values of
  the exponent~$\gamma$.} 
\label{fig.stretched}
\end{figure}

Equation~(\ref{eq.stretched}) is illustrated in
Fig.~\ref{fig.stretched} for different values of~$\gamma$ in two
different ways. Graph (a) (log-log) shows that decreasing the value
of~$\gamma$ the distribution starts from the exponential (Poisson)
case ($\gamma=1$) and approaches a power-law ($\gamma \rightarrow 0$)
with an exponent~$\alpha=1.5$. Graph (b) shows the distribution in the
form that the stretched exponentials are seen as straight
lines~\cite{bunde,bunde2,santhanam}. Generally, to obtain graph (b) from (a) one
needs to divide the distribution~$P(T)$ by the correct
pre-factor~$a=P(0)$, which is typically unknown.
Distribution~(\ref{eq.stretched}) shows the dependence of the
pre-factor~$a$ on the exponent~$\gamma$ when the stretched
exponential~$p_{\gamma}(\tau)$ is valid in the whole interval of
times. For experimental or numerical data, where neither~$a$
nor~$\gamma$ are known a priori, the relation between both is useful
to correctly visualize and fit the RTS. We note that in practice the
numerical fitting of the exponent~$\gamma$ is very sensitive and
typically depends on the choice of the pre-factor~$a$.


\subsection{\label{ssec.numerical} Numerical results for long-term correlated
  linear time series}

We compare now the stretched exponential
distribution~(\ref{eq.stretched}) to the numerical results of the RTS obtained
in long-term correlated 
time series. As in Ref.~\cite{bunde}, the data were generated using the
Fourier transform technique~\cite{prakash}: imposing a power-law decay on
the Fourier spectrum
\begin{equation}\label{eq.beta}
f_x(k) \propto k^{-\beta}\;,
\end{equation}
with $0<\beta<0.5$ and choosing phase angles at random, 
we obtain through an inverse Fourier transform the
long-term correlated time 
series in~$x$ with $\gamma_c=1-2\beta$ in
Eq.~(\ref{eq.correlation}). 
The data are Gaussian distributed with~$\langle x\rangle=0,\sigma=1$, and
Eq.~(\ref{eq.kac}) was used to calculate the
times $\tau=T/\langle T \rangle$.  

Having specified the power spectrum or, correspondingly, the
autocorrelation function for sequences of Gaussian random numbers
means to have fixed all parameters of a linear stochastic
process. Hence, in principle, the coefficients of an auto regressive (AR($r$))
or moving average (MA($r$)) process can be uniquely determined, where, due to
the power-law nature 
of spectrum and autocorrelation function, the orders $r$ of either of
these models have to be infinite~\cite{box}. Hence, the
following results are valid for the class of {\sl linear} long-term
correlated processes~\cite{bunde.prl}. In other words, higher order correlations for
this class of processes follow trivially from the two-point
correlations. 

\begin{figure}[!ht]
\centerline{\includegraphics[width=\columnwidth]{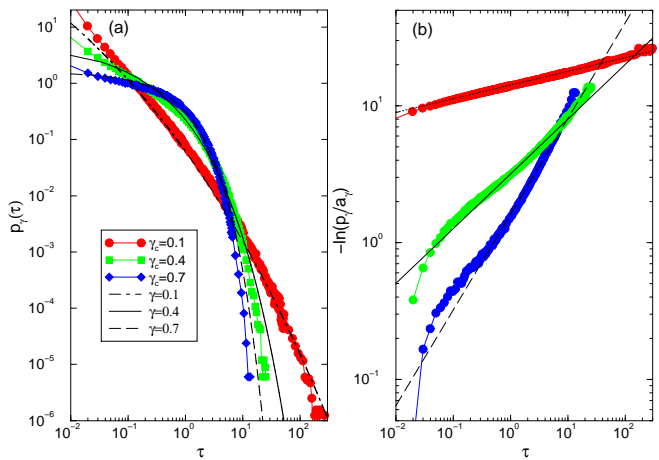}}
\caption{ (color online) RTS of long-term correlated linear
  time series with $N=2^{25} \approx 3\;10^7$ points
  for different values of~$\gamma_c$ (symbols). Lines are the stretched
  exponential distribution~(\ref{eq.stretched}) 
  with $\gamma=\gamma_c$. The recurrence interval is extreme with~$\mu(I_{ext})=10^{-2}$.}
\label{fig.gammas}
\end{figure}

We show in Fig.~\ref{fig.gammas} that the
stretched exponential distribution~(\ref{eq.stretched}) with~$\gamma=\gamma_c$
describes considerably well the RTS, obtained using extreme intervals
(Eq.~(\ref{eq.q})), of long-term correlated linear time series. The
agreement is 
especially good for small values of~$\gamma_c$ (long correlations)
and~$q\rightarrow \infty$ (which is equivalent to~$\mu(I)\rightarrow
0$). This result is a generalization of the result~{(\em ii)}~\cite{bunde}
since, using Eq.~(\ref{eq.stretched}) and considering~$\gamma=\gamma_c$, the
comparison between the theoretical and numerical distributions has no free
parameter and no fitting is made.  

\begin{widetext}

\begin{figure}[!ht]
\centerline{\includegraphics[width=14cm]{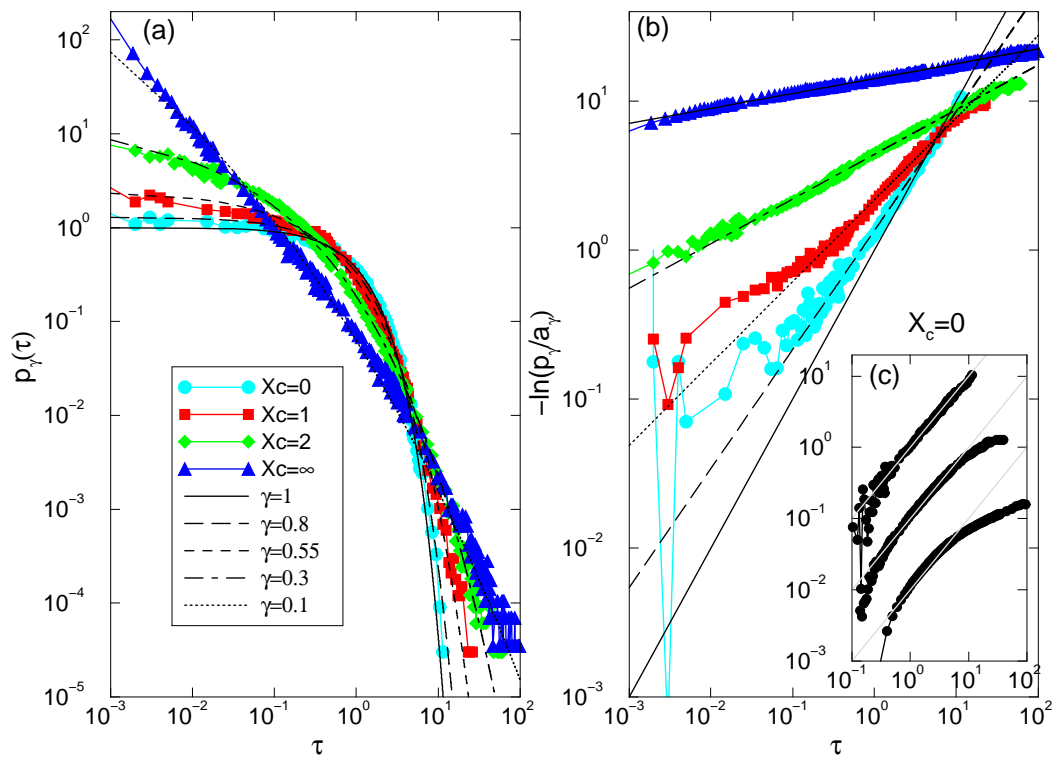}}
\caption{ (color online) RTS of long-term correlated linear time
  series with~$\gamma_c=0.1$ and different recurrence intervals (centered in~$X_c$ with
  measure~$\mu(I)=10^{-3}$). The lines are stretched exponentials
  distributions and the symbols connected by lines are the numerical
  simulations. From (a) to (b) we use the values given by $a_\gamma$ of the
  best fitting of Eq.~(\ref{eq.stretched}) in~(a). In (c) we analyze the case~$X_c=0$ for different
  values of $\mu(I)$, from bottom to top: $\mu(I)=10^{-1}$ (shifted down by
  $10^2$), $\mu(I)=10^{-2}$ (shifted by $10$) and $\mu(I)=10^{-3}$. The gray lines
are the Poissonian distribution ($\gamma=1$ in
  Eq.~(\ref{eq.stretched})). $a_{\gamma=1}=1$ was used for all three cases.}
\label{fig.Xc}
\end{figure}

\end{widetext}

Furthermore, we verify in Fig.~\ref{fig.Xc} that, for small~$\mu(I)$,  the
distribution~(\ref{eq.stretched}) is valid also for recurrence intervals
in the inner part of the data range (centered at $X_c$ and defined by
Eq.~(\ref{eq.xc})).  
When~$X_c\rightarrow\infty$, approaching the extreme interval, the
value of~$\gamma$ in Eq.~(\ref{eq.stretched}) approaches the value of the
correlation exponent~$\gamma_c$. Decreasing the value of~$X_c$ towards the
mean value of the PDF ($\langle x \rangle =0$) results in an increase
of~$\gamma$. This case was analyzed carefully in Fig.~\ref{fig.Xc}c, where the
dependence of  
the RTS on the size of the recurrence interval~$\mu(I)$ is shown. While for big
intervals the stretched exponential seems not to hold, when~$\mu(I)
\rightarrow 0$ (the limit Poincar\'e was interested in) the
distribution for $X_c=0$ tends to the upper limit~$\gamma=1$, the Poisson
distribution. 

In summary, the RTS of long-term correlated linear
time series with exponent $\gamma_c$, in the
limit of small interval~$\mu(I)\rightarrow 0$, is described by the
stretched exponential distribution (Eq.~(\ref{eq.stretched})) for all
recurrence times~$T$ and for recurrence intervals of both types~(\ref{eq.q})
and~(\ref{eq.xc}). The exponent~$\gamma$ is a continuous and monotonically 
decreasing function of the center~$X_c$ of the recurrence interval, with the limits
\begin{equation}\label{eq.limits}
\gamma = \left\{ \begin{array}{ll}
                  \gamma_c & \mbox{when } X_c\rightarrow \infty \mbox{ (extreme)},\\
		  1 & \mbox{when } X_c = 0 .
\end{array}
\right.
\end{equation}
This result has a simple interpretation in terms of the long-term correlations
contained in the time series. Calculating the RTS to
a specific 
interval measures the correlation between events inside this interval.
In this sense, our result suggests that the long-term 
correlations of the time series are concentrated in the extreme events (large
fluctuations) and vanish for events near the mean value (small
fluctuations). Relation~(\ref{eq.limits}) can then be interpreted as: 
approaching pure  extreme events ($\mu(I)\rightarrow 0$ and
$X_c\rightarrow\infty$) the RTS shows the whole correlation and
thus~$\gamma=\gamma_c$. Approaching pure middle events
$(\mu(I)\rightarrow0$ and $X_c=0$) no correlation is detected and consequently
the Poisson distribution $(\gamma=1)$ is recovered. 


\subsection{Change of observables}\label{ssec.observables}

The link between recurrence times on time series and Poincar\'e
recurrences (Sec.~\ref{ssec.poincare}) motivates the issue of the change of observables. All of
the empirical data exhibiting long-term correlations mentioned before
represent systems which involve a huge number of degrees of
freedom. Hence, there is a similarly huge arbitrariness in choosing a
given observation function $x(\vec\gamma)$, and the natural question
is what to expect when we change this observation function.

For instance, the correlations in the weather can be studied
through records of the daily maximum temperature or of the daily
precipitation~\cite{bunde2}. For the first observable, long-term correlations
for times larger than~$10$ days were 
found with an exponent~$\gamma \approx 0.7$ for continental
stations, independent of the location and of the climatic zone of the weather
station. On the other hand, the series of precipitation, 
obtained in the same locations and
for the same time windows, are not long-term correlated. A similar situation
is observed in financial market data. While the fluctuation of prices are
typically uncorrelated the volatility is long-term correlated~\cite{bouchaud}.
This gives already a clue that correlations measured on a given time
series do in fact characterize the fluctuations of the given
observable but do not characterize the underlying system in a more
abstract way. 

\begin{figure}[!ht] 
\centerline{\includegraphics[width=\columnwidth]{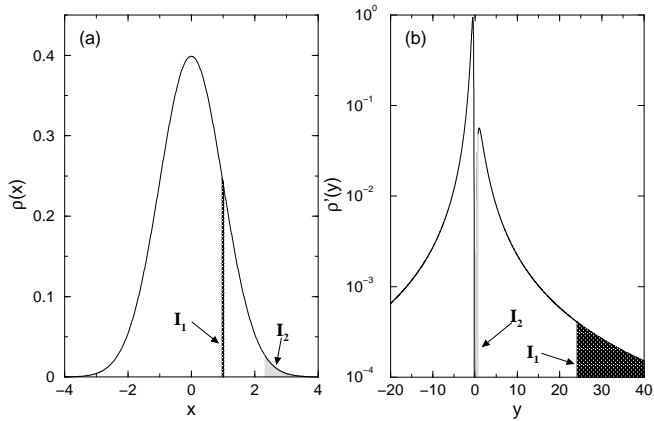}}
\caption{PDF of the series (a) of $x$ (Gaussian) and (b) of $y$
  (Eq.~(\ref{eq.pdfy}) with~$X_c=1,\delta=0.0207$). The points inside the interval~$I_1(X_c=1,\delta=0.0207)$ in $x$ become
  extreme events in $y$. The opposite happens for the extreme interval~$I_2=I_{ext}(q=2.3263)$.}
\label{fig.pdf}
\end{figure}

Here we want to study the dependence of correlations and RTS
on the chosen observable in more detail by comparing the properties
of different observables.
Generally, both observables~$x$ and~$y$ are functions
of the $d-$dimensional phase space vectors~$\vec{\gamma}$, and no simple function
connecting $x$ and $y$ exists.
Since we are starting from time series data without underlying
multi-dimensional phase space, we will restrict the analysis to a
subclass of changes of observables, where in fact $y$ is given by a
nonlinear (potentially non-invertible) function of $x$.
Hence, we construct time series of different observables~$y$ as
functions of the original long-term correlated time 
series of the variable~$x$. Having in
mind a recurrence interval defined through~($X_c,\delta$) by 
relation~(\ref{eq.xc}), consider the following reversible transformation
\begin{equation}\label{eq.y}
y_n=\frac{1}{x_n-(X_c-\delta)},
\end{equation}
which is essentially the inverse of the original series~$\{x_n\}$.  If the $x$-series
is Gaussian distributed as considered previously, the
PDF of the new series~$\{y_n\}$ is given by
\begin{equation}\label{eq.pdfy}
\rho'(y)=\frac{1}{\sqrt{2\pi}} \frac{1}{y^2} e^{-\frac{1}{2}(1/y+(X_c-\delta))^2},
\end{equation}
which is illustrated in Fig.~\ref{fig.pdf} for the
case~$X_c=1,\delta=0.0207$. In this figure it is also shown that the interval~$I_1$,
defined by the same~($X_c,\delta$) in~$x$, is transformed into an extreme interval
in~$y$. On the other hand, the extreme 
interval~$I_2$ in~$x$ is transformed into a recurrence interval in the middle of
the PDF of~$y$.
Since the
sequence of recurrence times~$T$ obtained using the original intervals in
the~$x$-series is also obtained using the transformed intervals in the
$y$-series, the RTS remains invariant under simultaneous
transformation of variables and recurrence intervals. Therefore, the
previous observation that the change of the recurrence interval in the
$x$-series does not affect the functional form of the stretched
exponential distribution~(\ref{eq.stretched}), but does affect the exponent $\gamma$, carries over to
transformations of the form~(\ref{eq.y}).
For instance, the RTS of a series obtained from
transformation~(\ref{eq.y}) applied to a time series~$x$ with~$\gamma_c=0.1$,
is well described by the stretched exponential distribution~(\ref{eq.stretched}) with (see
Fig.~\ref{fig.Xc}): $\gamma\approx0.55$ for extreme 
interval ($I_1$ in Fig.~\ref{fig.pdf}b) and~$\gamma=0.1$ for central interval
($I_2$ in Fig.~\ref{fig.pdf}). This result holds for all reversible
transformations.

\begin{figure}[!ht]
\centerline{\includegraphics[width=\columnwidth]{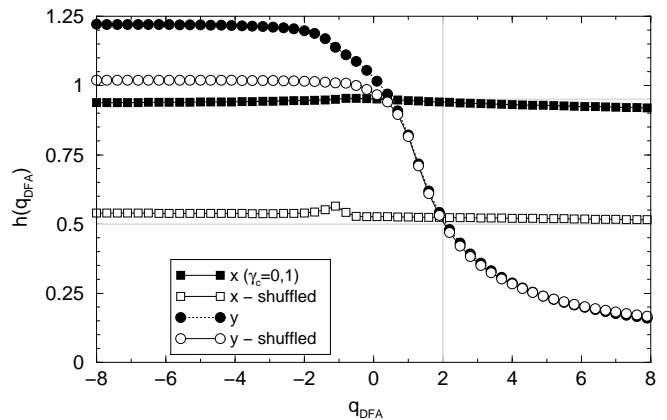}}
\caption{Generalized Hurst exponent of the time series of~$x$ and~$y$
  ($N=2^{20}\approx 10^6$ points) as a function of the scale~$q_{DFA}$. The
  horizontal gray lines are the non-correlated value ($h=0.5$) and the
  expected value for $\gamma_c=0.1$ ($h=1-\gamma_c/2=0.95$). The difference
  between the original and the shuffled time series measures the effect of the
  correlation at each scale~$q_{DFA}$.}
\label{fig.mf.gammas}
\end{figure}

An important fundamental question in this context is the behavior of the
long-term correlations under transformations of variables. Whereas 
the normalized autocorrelation function remains unchanged 
under shifts and rescalings of $x$, 
this is not the case under
transformations like~(\ref{eq.y}), where the transformed time series
of~$y$ is not long-term correlated at all, despite the long-term
correlations of the original $x$-series (see Fig.~\ref{fig.mf.gammas}
where $h(2)\approx 0.5$).  We characterize the $y$-series using the
multi-fractal detrended fluctuation analysis~\cite{kantelhardt}, which
is a much more powerful tool than the simple autocorrelation function,
since for different values of the parameter~$q_{DFA}$ different scales of
fluctuations are amplified.  In order to distinguish between the
multifractality due to long-term correlations and due to a broad PDF,
the typical procedure is to shuffle the time series randomly, i.e., we choose
randomly a new order of the~$N$ points of the original time series. Since the shuffled 
series loses all its temporal correlations but retains the same
PDF, the difference between the results of the two series (original
and shuffled) is exclusively due to temporal correlations.  In
Fig.~\ref{fig.mf.gammas} we show the multi-fractal analysis
(MF-DFA1~\cite{kantelhardt}) for the long-term correlated, Gaussian
distributed, linear time series~$\{x_n\}$ and for the transformed (through
Eq.~(\ref{eq.y})) time series~$\{y_n\}$. As expected, in the first case
roughly a single generalized Hurst exponent~$h(s)$ is obtained for all
scales in the original ($h(s)=1-\gamma_c/2=0.95$) and
shuffled~($h(s)=0.5$) time series. Due to the broad tails present in
Eq.~(\ref{eq.pdfy}), both the $y$-series and its 
shuffled version have multi-fractal spectrum, shown by the nontrivial
dependence of~$h(q_{DFA})$ on~$q_{DFA}$. The difference between the two, which
measures the effect of the temporal correlations, appears for small
scales, where the generalized Hurst exponent of the shuffled series is
smaller. This result is consistent with the interpretation made at the
end of Sec.~\ref{ssec.numerical} that the correlations of the
$x$-series is concentrated on the extreme events. Through
transformation~(\ref{eq.y}), the extreme events in~$x$ are mapped into very
small fluctuations in~$y$ and the temporal correlations of~$\{y_n\}$ are coherently noticeable for small
values of~$q_{DFA}$.

Through transformation~(\ref{eq.y}) we provide an example of
equivalence between the RTS 
of different observables obtained using extreme intervals, and 
the RTS calculated in the same series but using different recurrence
intervals. Always when the transformation of observables is invertible, there
exist a one-to-one correspondence between the original extreme values and a
new interval. This provides
another justification to the generalization of the recurrence of extreme
events to general recurrence intervals, proposed in
Sec.~\ref{ssec.poincare} inspired by the 
analogy to the Poincar\'e recurrences.

\subsection{The series of recurrence times}\label{ssec.seriesT}

It is also interesting to apply the distinction between the time
properties of the series and its PDF, discussed in
Sec.~\ref{sec.II}, to the series of recurrence times
$\{T_1,T_2,...T_{Ne}\}$ itself~\cite{multifractal}. In this case this means that the PDF,
which is the RTS of the original time series, is independent of its
correlation and shows that the results~{\em(ii)} and~{\em(iii)} stated
in Sec.~\ref{ssec.bunde} are independent. 
This is an important remark when prediction algorithms are considered, since in
many cases the correlation between the waiting times is more important
than their distribution~\cite{mega.prl}.

\begin{figure}[!ht]
\centerline{\includegraphics[width=\columnwidth]{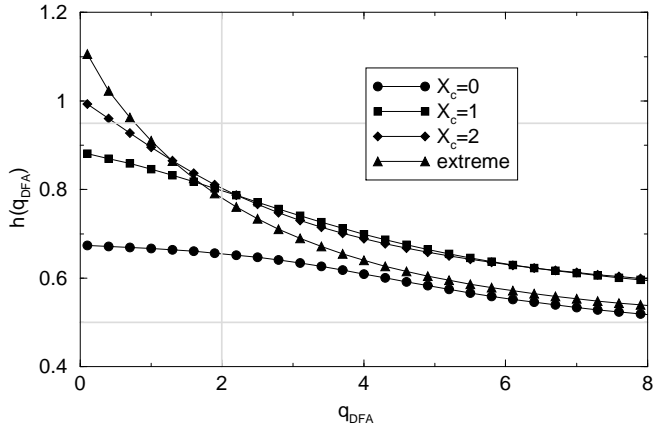}}
\caption{Multifractal spectrum for the series of recurrence
  times~$(\{T_1,T_1,...,T_{Ne}\})$ obtained for intervals with~$\mu(I)=0.01$, and
  different values of~$X_c$ in long-term correlated linear time series
  with~$\gamma_c=0.1$ ($N_e = 2^{18} \approx 2\;\times10^5$).}   
\label{fig.mf.serie}
\end{figure}

 The result~{\em (iii)} of Ref.~\cite{bunde} is
verified in Fig.~\ref{fig.mf.serie} through the multifractal analysis
of the series of  
recurrence times~$T$. Instead of the same correlation exponent we find a
multifractal spectrum. It is necessarily originated by the long-term
correlations since the PDF of these series are stretched exponential 
distributions, as verified in Fig.~\ref{fig.Xc}, which do
not have fat tails.

\section{\label{sec.IV} Discussion and Conclusion}

Well established regimes of decay of the RTS are exponentials
and power-laws, and, only recently observed, stretched exponentials. 
We have obtained a closed expression for the stretched exponential 
distribution of recurrence
times uniquely defined by the exponent~$\gamma$. As limits  $\gamma=1$
and $\gamma=0$, respectively, we recover the exponential and the
power-law decay from these (with the restriction that the power is
fixed to 3/2), suggesting that stretched exponentials describe recurrences in
systems that have neither exponential nor power-law RTS but that lie
in between these two cases. We have verified numerically that the
stretched exponential distribution is in good agreement with the
numerical results obtained for long-term correlated linear time
series, similarly to what was done in Ref.~\cite{bunde}. From the
point of view of these previous results, listed in
Sec.~\ref{ssec.bunde}, we have identified {\em(i)} with Kac's lemma;
generalized {\em(ii)} to the stretched exponential
distribution~(\ref{eq.stretched}), which is a function of a single
parameter and is valid for all recurrence intervals through
Eq.~(\ref{eq.limits}); and generalized {\em(iii)}, showing that the
sequence of recurrence times has a multi-fractal spectrum, with an
exponent~$\gamma_T$ different from~$\gamma_c$.  In order to verify if
the fluctuations around the stretched exponential distribution, shown
in the figures of Sec.~\ref{ssec.numerical}, are a consequence of
numerical limitations or real deviations, an analytical deduction of
the stretched exponential distribution~(\ref{eq.stretched}) is
necessary, what remains an open task.

Performing simple reversible transformations (like Eq.~(\ref{eq.y})) on the
original long-term correlated linear time series~$\{x_n\}$, we have shown that the
stretched exponential distribution characterizes also the RTS of extreme
events in time series that are not long-term correlated. 
The presence and absence of long-term correlations in the series of the
original observable~$x$ and of the transformed observable~$y$, respectively,
is similar to the one reported above for climatic records (temperature and  
precipitation) and stock-market indexes (volatility and fluctuation of
price). It is remarkable that this  
interesting behavior is obtained already through the simplest possible
approach, i.e., two different observables that depend directly and exclusively
on each other. These considerations
emphasize that the temporal characterization of the system through the
autocorrelation or RTS depend crucially on the
chosen observable.
By analyzing both the dependence of the exponent~$\gamma$ of the 
stretched exponential distribution with the center of the recurrence interval
(relation~(\ref{eq.limits})) and the multi-fractal spectrum of the $y$-series
(FIG.~\ref{fig.mf.gammas}) we conclude that, in long-term 
correlated linear time series, the correlations are concentrated in the
extreme events. 

Many interesting questions arise if one supposes that the measurements in a
given experiment lead to the time series of the observable~$y$,
introduced in Sec.~\ref{ssec.observables}, and that no natural
access to the observable~$x$ exists. The $y$-series has
a complex multi-fractal spectrum (Fig.~\ref{fig.mf.gammas}), a strange PDF
(Eq.~(\ref{eq.pdfy})) and a non-trivial dependence of the RTS with the 
recurrence interval. Nevertheless, through a simple transformation (the inverse
of relation~(\ref{eq.y})) one arrives at the $x$-series, that has a
mono-fractal spectrum, is Gaussian distributed and has a simple
(Eq.~(\ref{eq.limits})) dependence of the RTS on the recurrence interval. This
suggests the existence, in some situations, of 	``distinguished
observables'' where the time series analysis is extremely simplified. It is
an interesting open problem to develop a procedure able to determine the
transformation (when it exists) that lead to the           
``distinguished observables''. 


\begin{acknowledgments} 
The authors thank J. Davidsen for helpful discussions and for the careful reading
of the paper. E.G.A. thanks E.C. da
Silva and I.L.Caldas for illuminating discussions on related topics. 
This work was supported by CAPES (Brazil) and DAAD (Germany).
\end{acknowledgments}



\begin{thebibliography}{10}

\bibitem{bunde}
A.~Bunde, J.~F. Eichner, S.~Havlin, and J.~W. Kantelhardt.
\newblock {\em Physica A}, 330:1, 2003.

\bibitem{bunde.prl} 
A.~Bunde, J.~F. Eichner, J.~W. Kantelhardt and S.~Havlin.
\newblock {\em Phys. Rev. Lett.}, 94, 048701 (2005). 

\bibitem{bunde2}
A.~Bunde, J.~Eichner, R.~Govindan, S.~Havlin, E.~Koscielny-Bunde, D.~Rybski,
  and D.~Vjushin.
\newblock In {\em Nonextensive Entropy-Interdisciplinary
  Applications}. (Oxford Univ. Press, New York, 2003).
\newblock arXiv:physics/0208019.

\bibitem{santhanam}
M.~S. Santhanam and H.~Kantz.
\newblock {\em Physica A}, 345: 713, 2005.

\bibitem{alley}
To associate the recurrence time with residence time in stochastic resonances see
\newblock R.~B. Alley, S.~Anandakrishnan, and P.~Jung.
\newblock {\em Paleoceanography}, 16(2):190, 2001.


\bibitem{earthquakes}
P.~Bak et~al.
\newblock {\em Phys. Rev. Lett.}, 88(17):178501, 2002;
\newblock A.~Corral.
\newblock {\em Phys. Rev. Lett.}, 92(10):108501, 2004;
\newblock N.~Scafetta and B.~J. West.
\newblock {\em Phys. Rev. Lett.}, 92(13):138501, 2004;
\newblock J. Davidsen and C.~Goltz
\newblock {\em Geophys. Res. Lett.}, 31:L21612, 2004.

\bibitem{boffetta.prl}
G.~Boffetta et~al.
\newblock {\em Phys. Rev. Lett.}, 83(22):4662, 1999.

\bibitem{joern}
J.~Davidsen and H.~G. Schuster.
\newblock {\em Phys. Rev. E}, 65:026120, 2002.


\bibitem{murilo.plasma}
M.~S. Baptista, I.~Caldas, M.~Heller, and A.~A. Ferreira.
\newblock {\em Physics of Plasmas}, 8:4455, 2001.

\bibitem{murilo.bolsa}
M.~S. Baptista and I.~L. Caldas.
\newblock {\em Physica A}, 312:539, 2002.

\bibitem{altmann}
E.~G. Altmann, E.~C. da~Silva, and I.~L. Caldas.
\newblock {\em Chaos}, 14(4):975, 2004.


\bibitem{kac}
M.~Kac.
\newblock {\em Bulletin of the American Mathematical Society}, 53:1002, 1947.

\bibitem{zas.pr}
G.~M. Zaslavsky.
\newblock {\em Physics Reports}, 371:461, 2002.




\bibitem{sornette}
A.~Sornette and D.~Sornette.
\newblock {\em Europhys. Lett.}, 9(3):197, 1989.

\bibitem{yang.prl}
X.~Yang, S.~Du, and J. Ma.
\newblock {\em Phys. Rev. Lett.}, 92(22):228501, 2004.



\bibitem{christiansen}
K.~Christiansen and Z.~Olami.
\newblock {\em J. Geophys. Res.}, 97(B6):8729, 1992.

\bibitem{sanchez.prl}
R.~S\'anchez, D.~E.~Newman, and B.~A. Carreras.
\newblock {\em Phys. Rev. Lett.}, 88(6):068302, 2002.


\bibitem{hirata}
M. Hirata.
\newblock {\em Ergod. Th. Dyn. Systems}, 13(3):533,1993.
\newblock {Hirata et al.}
\newblock {\em Comm. Math. Phys.}, 206:33,1999.

\bibitem{kantelhardt}
J.~W. Kantelhardt, S.~A. Zschiegner, E.~Koscielny-Bunde, S.~Havlin, A.~Bunde,
  and H.~E. Stansley.
\newblock {\em Physica A}, 316:87, 2002.


\bibitem{thurner}
S. Thurner~et al.
\newblock {\em Fractals}, 5(4):565, 1997.
\newblock arXiv:adap-org/9709006.


\bibitem{prakash}
S.~Prakash, S.~Havlin, M.~Schwartz, and H.~E. Stanley.
\newblock {\em Phys. Rev. A}, 46(4):R1724, (1992).

\bibitem{box}
G.~E.~P. Box, G.~M. Jenkins, and G.~C. Reinsel.
\newblock
{\em Time series analysis : forecasting and control} (Prentice Hall, New Jersey, 1994).


\bibitem{joern2}
J.~Davidsen and M.~Paczuski.
\newblock {\em Phys. Rev. E}, 66:050101(R), 2002.

\bibitem{bouchaud}
M.~Potters, R.~Cont, and J.~Bouchaud.
\newblock {\em Europhys. Lett.}, 41(3):239, 1998.

\bibitem{multifractal}
The (multi-) fractal properties of the recurrence times were studied from different perspectives in: 
\newblock  V. Afraimovich and G. M. Zaslavsky, {\em Phys. Rev. E}, 55 (5):5418
(1997) and N.~Hadyn et~al. {\em Phys. Rev. Lett.}, 88 (22):224502 (2002). 

\bibitem{mega.prl}
M.~S. Mega et~al.
\newblock {\em Phys. Rev. Lett.}, 90(18):188501, 2003.







\end{thebibliography}
\end{document}